\documentclass[prc,aps,amsfonts,amsmath,superscriptaddress,floatfix,twocolumn]{revtex4}
\usepackage{graphicx,float} 
\usepackage{epsfig} 
\usepackage{rotating}
\usepackage{mwe}    
\usepackage{subfig}
\usepackage{xfrac}
\usepackage{xcolor}
\begin{document}

\title{Alpha-particle formation and clustering in nuclei}

\author{E. Khan}
\affiliation{IJCLab, Universit\'e Paris-Saclay, CNRS/IN2P3, 91405 Orsay Cedex, France}
\author{L. Heitz}
\affiliation{IJCLab, Universit\'e Paris-Saclay, CNRS/IN2P3, 91405 Orsay Cedex, France}
\author{F. Mercier}
\affiliation{IJCLab, Universit\'e Paris-Saclay, CNRS/IN2P3, 91405 Orsay Cedex, France}
\author{J.-P. Ebran}
\affiliation{CEA,DAM,DIF, F-91297 Arpajon, France}
\affiliation{Universit\'e Paris-Saclay, CEA, Laboratoire Mati\`ere en Conditions Extr\^emes, 91680, Bruy\`eres-le-Ch\^atel, France}

\begin{abstract} 
The nucleonic localization function has been used for a decade to study the formation of alpha-particles in nuclei, by providing a measure of having nucleons of a given spin in a single place. However, differences in interpretation remain, compared to the nucleonic density of the nucleus. In order to better understand the respective role of the nucleonic localization function and the densities in the alpha-particle formation in cluster states or in alpha-decay mechanism, both an analytic approximation and microscopic calculations, using energy density functionals, are undertaken. The nucleonic localization function is shown to measure the anti-centrifugal effect, and is not sensitive to the level of compactness of the alpha-particle itself. It probes the purity of the spatial overlap of four nucleons in the four possible (spin, isospin) states. The density provides, in addition, information on the compactness of an alpha-particle cluster. The respective roles of the nucleonic localization function and the density are also analyzed in the case of alpha-particle emission. More generally, criteria to assess the prediction of alpha-cluster in nuclear states are provided. 
\end{abstract}
 


\date{\today}

\maketitle

\section{Introduction}

The question of the formation of alpha-clusters in nuclei is an important issue for nuclear physics, as illustrated by the famous Hoyle state, or the study of the behavior of the nuclear fluid at low-density \cite{freer,beck,typ10,sch13,fre18,ebr20}. In principle, a sound indicator for alpha clusterization should be based on a four particles correlator measurement. However, such beyond mean-field calculations are expected to be heavy, especially in deformed nuclear states, where clusterization could occur. 

For a decade, the nucleonic localization function (NLF) has increasingly been used to tag the formation of alpha-particles \cite{rei11,zha16,sch17,naza2,ebr18,uma21,ren22}, such as in cluster states, or during the fission process. It is therefore important to understand its usefulness and limitations in details. Concomitantly, the use of Relativistic Mean Field (RMF) Energy Density Functional (EDF) approaches has been shown to pinpoint localized alpha structures in nuclei \cite{aru05,ebr12,ebr13,ebr14,ebr14a,zha15,zho16,mar18,mar19}. This is less the case with Skyrme EDF approaches, where localized alpha-particle structures do not appear in the density, unless strong constraints are considered, such as in the case of high-spin rotations \cite{ich11} or very large deformations \cite{rei11}. This difference in behavior between the RMF and Skyrme densities was related to the difference in the depth of the mean-field potential \cite{ebr12}.

However, some questions remain: are the RMF density and the NLF probing the same aspects of alpha-particle formation in nuclei ? Why do the alpha-particles indicated by NLF and RMF densities are not exactly located at the same position ? Why is the NLF extending very far outside the nucleus, typically a few times its radius ? How to explain that similar NLFs are found when using different EDF such as the Skyrme and the RMF one, but different densities are obtained (the RMF showing more alpha-particle formation than the Skyrme one) ? Is the NLF well designed to study clustering in nuclei ? An especially relevant case to study is the alpha emission process, which was recently described using relativistic EDF approaches \cite{mer20,mer21}. The investigation of both densities and NLF during the formation of the alpha-particle, just before its emission, shall enlighten their respective roles, and provide interesting information on the question of the alpha-preformation process in nuclei, before their radioactive decay. 

The present work intends to help in clarifying these questions, in order to provide a sound procedure for analyzing the presence of alpha-particles in nuclei. The complementary roles of the NLF and the RMF density shall be investigated. In particular, the question of whether the NLF only can monitor the formation of alpha-particle clustering in nuclei shall be addressed. Section II provides an analytical study of the NLF, in order to understand the basic mechanism driving this quantity. Section III focuses on microscopic calculations of densities and NLF in nuclei, with an application to the case of the alpha decay of $^{212}$Po. Criteria to identify alpha-particle clusters in nuclear states are also discussed.

\section{Analytic calculation of the localization function}

The key ingredient of the NLF is the leading term $Z_{q\sigma}(\vec{r})$ of the Taylor expansion of the same-spin ($\sigma$) and same-isospin ($q$) conditional pair probability ~\cite{rei11}:
\begin{equation}
Z_{q\sigma}(\vec{r})\equiv\tau_{q\sigma}(\vec{r})\rho_{q\sigma}(\vec{r})-\frac{1}{4}\left[\vec{\nabla}\rho_{q\sigma}(\vec{r})\right]^2-\vec{j}_{q\sigma}^{\phantom{-}\hspace{-0.14cm}2}(\vec{r}),
\label{eq:z}
\end{equation}
where $\rho_{q\sigma},\tau_{q\sigma},\vec{\nabla}\rho_{q\sigma}$ and $\vec{j}_{q\sigma}$ are the nucleon density, kinetic energy density, density gradient and current density, respectively. In the present static and time-reversal symmetric case, the current density $\vec{j}_{q\sigma}$ vanishes.
After normalization by $\tau_{q\sigma}^\text{TF}\rho_{q\sigma}$, where $\tau_{q\sigma}^\text{TF}=\frac{3}{5}(6\pi^2)^{\sfrac{2}{3}}\rho_{q\sigma}^{\sfrac{5}{3}}\equiv \frac{\rho_{q\sigma}^{\sfrac{5}{3}}}{a}$ is the Thomas-Fermi kinetic energy density, the NLF reads as \cite{rei11}
 \begin{equation}
C_{q\sigma}(\vec{r})=\left[1+\left(\frac{aZ_{q\sigma}(\vec{r})}{\rho_{q\sigma}^{8/3}(\vec{r})}\right)^2\right]^{-1}.
\label{eq:cdef}
\end{equation} 
A low conditional pair probability, i.e. a high localization of a nucleon at a given position, translates into $Z_{q\sigma}(\vec{r})=0$ and $C_{q\sigma}(\vec{r})=1$. 
Therefore, a $C_{q\sigma}(\vec{r})=1$ value, in a N=Z nucleus, indicates at least a pure spatial overlap of the four possible (isospin,spin) nucleonic states 
(q$\sigma$)=(n$\uparrow$,n$\downarrow$,p$\uparrow$,n$\downarrow$). This case is called hereafter a pure 4-nucleons overlap. It should not be confused with an alpha-particle cluster, which requires, in addition, spatial compactness in a 4 particles bound state. This last feature has no a priori reason to be monitored by the NLF, which is calculated at the mean-field level: in this framework, the conditional pair probability (\ref{eq:z}) shall not be sensitive to the possible correlated behavior of the 4 nucleons, beyond Pauli or deformation effects. 

In order to provide a global understanding of the mechanisms at work in the NLF, we first analytically study this quantity.

\subsection{The spherical case}

\subsubsection{Analytic derivation of the localization function}

In a first approach, spherical symmetry is assumed to be preserved at the mean-field level. This shall enable a detailed analysis of the NLF.
N=Z nuclei are considered and the spin-orbit term is neglected.

Omitting in the following the spin and isospin degrees of freedom, the nucleon wavefunction is 
\begin{equation}
\Phi_{n,l,m}(\vec{r})=\varphi_{n,l}(r)Y_l^m(\theta,\phi)
\label{eq:wfdef}
\end{equation}

where $n$ and $\ell$ are the radial and orbital quantum numbers, respectively, and $m$ ranges from $-\ell$ to $\ell$. 

In this case, the density and kinetic energy densities read, for a given (spin,isospin) value, in agreement with \cite{vau72}, as 

\begin{equation}
\rho_{q\sigma}(r)=\sum_{nl} \frac{2l+1}{4\pi}\varphi_{nl}^2(r)
\end{equation}

\begin{equation}
\tau_{q\sigma}(r)=\sum_{nl} \frac{2l+1}{4\pi}\left[\left(\partial_r\varphi_{nl}(r)\right)^2+\frac{l(l+1)}{r^2}\varphi_{nl}(r)^2\right]
\label{eq:tau}
\end{equation}

As an aside, this expression for the kinetic energy density can be used to derive a relation involving the derivative of the spherical harmonics, not found in textbooks, to our knowledge (see App.~\ref{app}).

In the spherical symmetry case, using Eq. (\ref{eq:wfdef}) in Eq. (\ref{eq:z}), leads to

\begin{eqnarray}
\label{eq:zspher}
\begin{split}
Z_{q\sigma}(r)& =\sum_{nl,n'l'}\frac{(2l+1)(2l'+1)}{16\pi^2}[(\partial_r\varphi_{nl})^2\varphi_{n'l'}^2\\
&\quad +\frac{l(l+1)}{r^2}\varphi_{nl}^2\varphi_{n'l'}^2- \varphi_{nl}\varphi_{n'l'}(\partial_r\varphi_{nl})(\partial_r\varphi_{n'l'})]
\end{split}
\end{eqnarray}

It should be reminded that a pure 4-nucleons overlap condition is met for $C_{q\sigma}=1$, hence for $Z_{q\sigma}=0$. In order to further derive an analytic expression, the present derivation is applied to the case of 3D Harmonic Oscillator (HO) wavefunctions. In this case, the following relation is obtained:

\begin{equation}
\partial_r\varphi_{nl}=\varphi_{nl}\left(\frac{l}{r}-\frac{r}{b^2}\right)-\frac{2\sqrt{n-1}}{b}\varphi_{n-1,l+1}
\label{eq:hod}
\end{equation}

where $b$ is the oscillator length of the spherical 3D HO wavefunction.

Eqs. (\ref{eq:zspher}) and (\ref{eq:hod}) lead to the following expression for $Z_{q\sigma}$:

\begin{equation}
\label{eq:zfin}
\begin{split}
Z_{q\sigma}(r)&=\sum_{nl}\frac{(2l+1)^2}{16\pi^2}\frac{\varphi_{nl}^4}{r^2}l(l+1)
+\sum_{n'l'>nl}\frac{(2l+1)(2l'+1)}{16\pi^2}\\
&\quad\frac{\varphi_{nl}^2\varphi_{n'l'}^2}{r^2}\left[l(l+1)+l'(l'+1)+(l-l')^2\right]\\
&\quad+\sum_{n'l',nl}\frac{4\sqrt{n-1}}{br}(l'-l)\varphi_{nl}\varphi_{n-1,l+1}\varphi_{n'l'}^2\\
&\quad+\frac{4(n-1)}{b^2}\varphi_{n'l'}^2\varphi_{n-1,l+1}^2\\
&\quad-\frac{4\sqrt{(n-1)(n'-1)}}{b^2}\varphi_{nl}\varphi_{n'l'}\varphi_{n-1,l+1}\varphi_{n'-1,l'+1}
\end{split}
\end{equation}

It should be noted that in the case of light nuclei, where only $n=1$ states are filled, only the first two terms of $Z_{q\sigma}(r)$ remain. These terms are driven by the orbital angular momentum and originate from the centrifugal effect. Therefore, the smallest values of $Z_{q\sigma}(r)$ (i.e. pure 4-nucleons overlap) are expected on the surface of the nucleus. Moreover, in the case of the $^4$He, Eq. (\ref{eq:zfin}) gives $Z_{q\sigma}(r)=0$ over all the space, because only the 1s state is filled for each nucleon and spin type.

\subsubsection{Calculations on $^{4}$He and $^{16}$O}

Let us start with the case of $^4$He. As mentioned above, $Z_{q\sigma}(r)=0$. Hence, (\ref{eq:cdef}) gives $C_{q\sigma}=1$, and the NLF predicts a pure 4-nucleons overlap over the space. This is in agreement with the microscopic computation of the NLF using the Skyrme EDF, see Fig. 1 of \cite{rei11}. It provides a nice explanation for the NLF equals to 1, in the case of a mere alpha-particle: this comes from the $\ell=0$ value of the wavefunction of all its nucleons (Eq. (\ref{eq:zfin})).

However, Eq. (\ref{eq:cdef}) shows that outside of the nucleus, the $\rho^{16/3}$ quantity of the denominator tends toward 0, whereas the numerator is also here equal to zero. This issue impacts all the NLF calculations and will be addressed in the next subsection. 

In the case of $^{16}$O, both the 1s and 1p HO states are filled. Hence, the following expression is obtained for $Z_{q\sigma}$, using Eq. (\ref{eq:zfin}):
 
\begin{equation}
Z_{q\sigma}(r)=\frac{9}{16\pi^2}\frac{\varphi_{1p}^2(r)}{r^2}\left(2\varphi_{1p}^2(r)+\varphi_{1s}^2(r)\right)
\label{eq:z1s1p}
\end{equation}
 
One sees again the presence of a centrifugal $1/r^2$ term, which drives $Z_{q\sigma}$ towards small values when reaching the surface of the nucleus. Since small values of $Z_{q\sigma}$ correspond to values of $C_{q\sigma}$ close to 1 (see Eq. (\ref{eq:cdef})), the NLF could be considered as a probe of the anti-centrifugal effect. This leads to the interpretation that a small centrifugal effect favors the formation of a pure 4-nucleons overlap. This is in agreement with the above discussion on $C_{q\sigma}=1$ when only the $\ell=0$ state is filled ($^4$He).
 
\begin{figure}[tb]
\scalebox{0.35}{\includegraphics{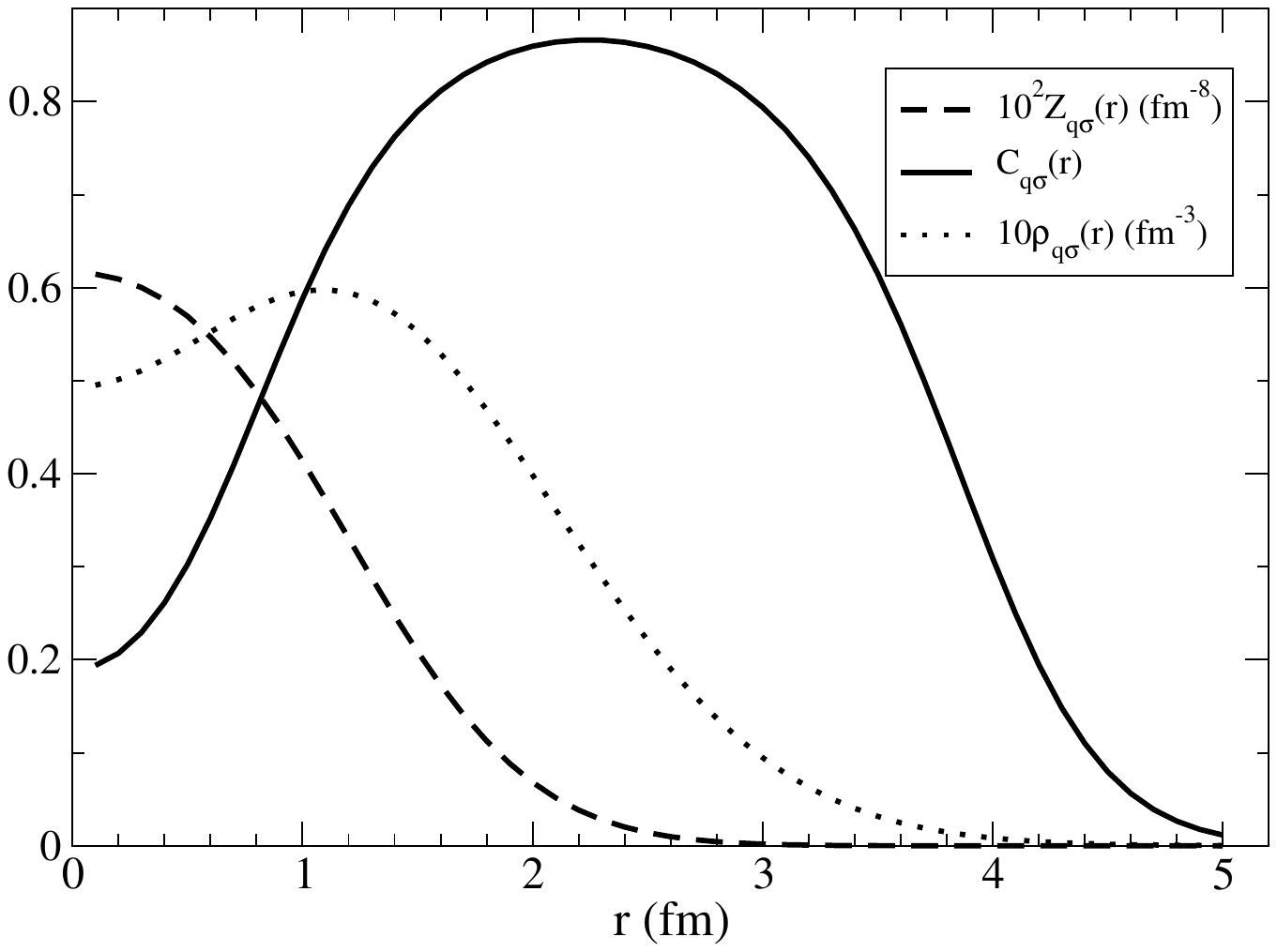}}
 \caption{Nucleonic density, localization function $C$ (Eq. (\ref{eq:cdef})) and $Z$ (Eq. (\ref{eq:z1s1p})) functions for $^{16}$O, calculated with the 3D HO approximation, for a given (spin,isospin) value.}
 \label{fig:16Oloc}
\end{figure}

Fig. \ref{fig:16Oloc} displays the behavior of the localization function in $^{16}$O, using Eqs. (\ref{eq:z1s1p}) and (\ref{eq:cdef}), as well as the behavior of $Z_{q\sigma}(r)$ and the nucleonic density. This result is in agreement with the one calculated microscopically (Fig. 2 of \cite{rei11}). It should be noted that the localization function is close to one around the surface of the nucleus, which seems to be a general mechanism. Of course, because of spherical symmetry, there is no formation of alpha cluster, but only a ring of larger probability to detect a pure 4-nucleons overlap around the surface of the nucleus. The key point is that the $Z_{q\sigma}$ function drops before the density (typically 2 fm before), because of the $1/r^2$ term of Eq. (\ref{eq:z1s1p}). In this region close to the surface of the nucleus, where $Z_{q\sigma}(r)$ is small and $\rho_{q\sigma}$(r) non-negligible, Eq. (\ref{eq:cdef}) shows that it implies C$_{q\sigma}\simeq$1. Therefore, the NLF reaches large values in this area, which is also in agreement with its anti-centrifugal interpretation.

\subsubsection{Interpretation with the density and localization parameters}

We shall further interpret the behavior of the localization function at the surface of the nucleus, and beyond, namely: i) the surface is a critical area for the NLF, where $C_{q\sigma}$ gets close to 1, and ii) the large spatial extension of the NLF beyond the surface. For this purpose, let us further approximate the value of $Z_{q\sigma}$(r), for this range around the surface of the nucleus, by considering the most contributing wavefunction on the surface, i.e. the one with the largest value of $\ell$, named hereafter $\ell_{max}$.

For $n_\text{max}=1$, which is realized for light and most of medium-mass nuclei, one gets for the main l.h.s term of Eq. (\ref{eq:cdef}):

\begin{equation}
\frac{aZ_{q\sigma}(r)}{\rho_{q\sigma}^{8/3}(r)}\simeq \frac{5}{3}\frac{\ell_{max}(\ell_{max}+1)}{(6\pi)^{2/3}}\frac{1}{r^2\rho_{q\sigma}^{2/3}}
 \label{eq:aZ}
\end{equation}

In order to provide an interpretation, let us consider the density parameter $r_0$, related to the density by

\begin{equation}
\rho_{q\sigma}=\frac{1}{4}\left(\frac{4}{3}\pi r_0^3\right)^{-1}
 \label{eq:rr}
\end{equation}

where the 1/4 factor comes from the spin and isospin degeneracies.

\begin{figure}[tb]
\scalebox{0.35}{\includegraphics{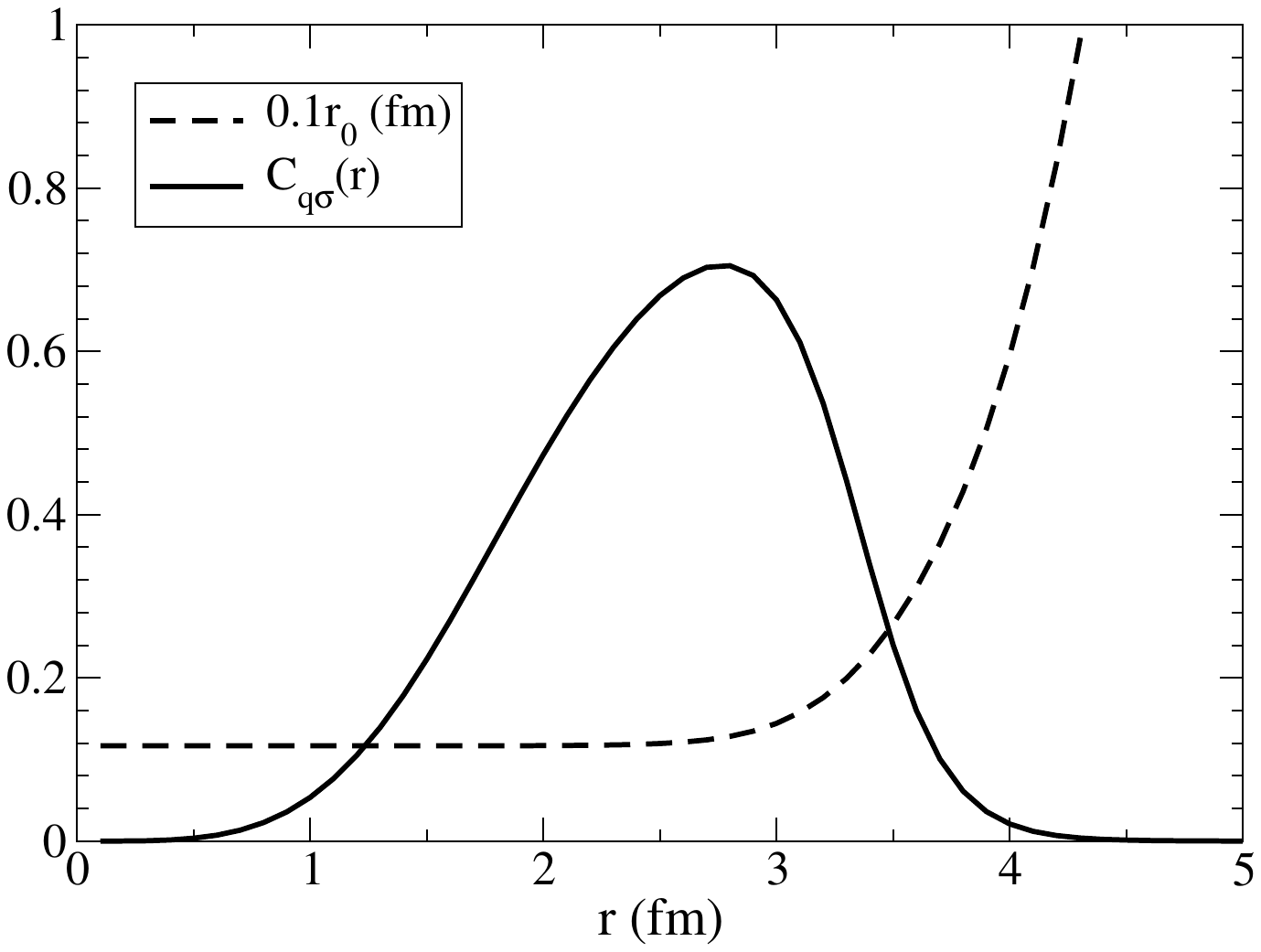}}
 \caption{Density parameter $r_0$ (\ref{eq:rr}) and NLF (\ref{eq:cdef}) calculated with the surface approximation (\ref{eq:Zsurf}), in the case of $^{16}$O.}
 \label{fig:16Or0}
\end{figure}

$r_0$ can be interpreted as the typical internucleon distance. At saturation density, one gets $r_0\simeq 1.2$ fm. On the surface of the nucleus, the density drops, and hence $r_0$ increases,  as shown on Fig. \ref{fig:16Or0}. 
Using (\ref{eq:rr}) in (\ref{eq:aZ}), one gets 

\begin{equation}
\frac{aZ_{q\sigma}(r)}{\rho_{q\sigma}^{8/3}(r)}\simeq \frac{5}{3}\left(\frac{8}{9}\right)^{2/3}\ell_{max}(\ell_{max}+1)\left(\frac{r_0}{r}\right)^2
 \label{eq:Zsurf}
\end{equation}

allowing to calculate the NLF with this approximated expression.

Fig. \ref{fig:16Or0} displays the behavior of $r_0$, using a Woods Saxon density for $\rho(r)$, and subsequently, the one of the NLF (from Eqs. (\ref{eq:cdef}) and (\ref{eq:Zsurf})) in the case of $^{16}$O ($\ell_{max}=1$), showing the typical bell shape on the surface, in agreement with the microscopic calculations \cite{rei11} and Fig. \ref{fig:16Oloc} of the present work. $r_0$ is constant in the nucleus, explaining why the NLF gets close to 1 at the surface, due to the drop of the $r_0/r$ ratio. Beyond the surface, $r_0$ starts to increase faster than $r$, implying a drop in the NLF value (see Eq. (\ref{eq:Zsurf})). 

This analysis shows that the dimensionless ratio $r_0/r$ drives the NLF. From Eq. (\ref{eq:Zsurf}), the lengths involved in the NLF are the position $r$ and the density parameter $r_0$, but there is no information on the spatial dispersion of the alpha-particle itself. In order to include this information, quantities such as the so-called localization parameter, $\alpha_{loc}$=$\Delta r/r_0$ \cite{ebr12,ebr13}, where $\Delta r$ is the dispersion of the nucleonic wavefunction, should be considered.  $\alpha_{loc}$ is a measure of the formation of localized (in the sense of not dispersed) alpha-particle. This last parameter is analogous to the Wigner or Brueckner parameter in condensed matter \cite{wig34}. Therefore, the NLF probes the purity of the spatial overlap of the 4 nucleons of the alpha-particle (n$\uparrow$,n$\downarrow$,p$\uparrow$,n$\downarrow$), irrespective of their own localization (non-dispersion), as illustrated by the present spherical case, where the 4 nucleons behave as a ring. It doesn't indicate whether these 4 nucleons form a bound and localized (compact) state or not: the four nucleons state, identified by the $C_{q\sigma}=1$ condition of the NLF, could be delocalized and/or unbound.

In the $C_{q\sigma}=1$ condition, as discussed above, the pure 4-nucleons overlap shall be focused on the surface of the nucleus, where the centrifugal effect is weak. Therefore, the NLF, (as a necessary C$_{q\sigma}$=1 condition) could be a first signal for an alpha condensation in low-density systems, such as the related Mott effect \cite{typ10,sch13,ebr20}. However, additional indicators are necessary, such as the localization parameter $\alpha_{loc}$, related to the formation of localized alpha-particles, and hence clusterization.

It should be noted that outside of the nucleus (typically beyond its radius, such as r $\sim$ 3 fm in the case of $^{16}$O), the nucleonic density is small ($\rho\lesssim 0.2\rho_0$, where 
$\rho_0$ is the saturation density), whereas the NLF can still have large values close to 1 (see Fig. \ref{fig:16Oloc}). This means that there can be a pure 4-nucleons overlap of the very tails of the four nucleonic wave functions in this region. However, the density integral there is of course much smaller than 4. Therefore, this case cannot correspond to alpha-particle clustering. This spatial region can be confusing, when only the NLF is considered in order to trace alpha-particle clustering. It is sometimes artificially cured by adding an arbitrary spatial cutoff \cite{zha16,sch17,naza2}. But as long as both the nucleonic density and the NLF are considered in order to look for alpha-particle clustering, there is no difficulty in interpretation, as it will be also discussed in section \ref{sec:micro}.

\subsection{The deformed case}

At the mean-field level, the proper occurrence of clustering requires a deformed nucleus \cite{freer}, emphasizing that C$_{q\sigma}$=1 is a necessary, but not sufficient condition for alpha clustering, as it can be reached in spherical nuclei (see present Fig. \ref{fig:16Oloc}, and Fig. 2 of Ref. \cite{rei11}). For this purpose, we extend the previous study to the case of the deformed HO, using the following definition of the wavefunction:

\begin{equation}
\label{eq:fodef}
\phi_{n_z,n_r,m_l}(r_\perp,z,\varphi)=\phi^{m_l}_{n_r}(r_\perp)\phi_{n_z}(z)\frac{e^{im_l\varphi}}{\sqrt{2\pi}}
\end{equation}

where the corresponding Nilsson quantum numbers are defined in \cite{rs80}.

It is relevant, in order to simplify the interpretations, to focus on the $r_\perp=0$ axis, since clusters are known to belong to this axis, in axial symmetry (see e.g. \cite{ebr14}). In this case, only $m_l=0$ states remain, and one gets from Eqs. (\ref{eq:z}) and (\ref{eq:fodef}):

\begin{equation}
\label{eq:zdef}
\begin{split}
Z(z)=\frac{2}{b^4_\perp b^2_z\pi^2}\sum_{n_z\neq n'_z}&\phi_{n'_z}\phi_{n_z-1}[n_z\phi_{n'_z}\phi_{n_z-1}\\
&\quad -\sqrt{n_zn'_z}\phi_{n_z}\phi_{n'_z-1}]
\end{split}
\end{equation}

where $b_{\perp,z}$ denotes the oscillator lengths along the respective coordinates.

Let us first consider the case of $^{8}$Be. The only $m_l=0$ states are [$N$ $n_z$ $m_l$ $\Omega$]=[000 1/2] and [110 1/2]. Eq. (\ref{eq:zdef}) becomes

\begin{equation}
Z(z)=\frac{2\phi_0^4}{\pi^2 b^4_\perp b_z^2}
\end{equation}

\begin{figure}[tb]
\scalebox{0.35}{\includegraphics{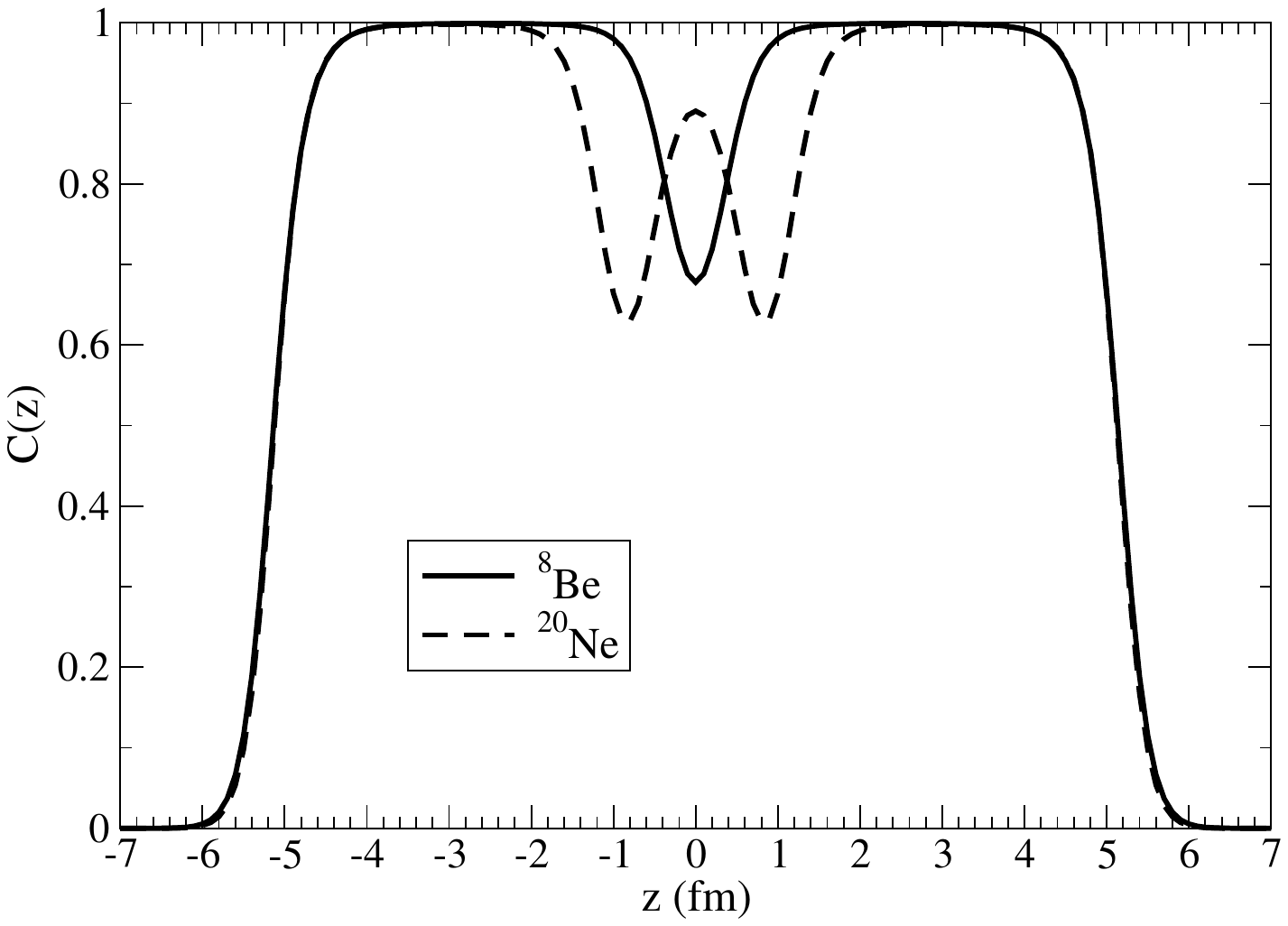}}
 \caption{NLF function using the deformed HO approach, for $r_\perp=0$ and $b_\perp=b_z$, for  $^8$Be and $^{20}$Ne.}
 \label{fig:cdef}
\end{figure}

Fig. \ref{fig:cdef} displays the corresponding NLF, showing a similar shape than in the case of $^{16}$O (Figs. \ref{fig:16Oloc} and \ref{fig:16Or0}).
Therefore, there is a general mechanism at work, leading to a peaked NLF at the surface of the nucleus. It should be noted that in the case of $^8$Be, due to axial symmetry, cluster structures can appear, contrary to the $^{16}$O case, where the spherical symmetry is not compatible with cluster formation.

In the case of $^{20}$Ne, the $m_l=0$ states are, in the prolate case: [000 1/2], [110 1/2] and [220 1/2]. This leads to the following expression of $Z$:

\begin{equation}
Z(z)=\frac{2}{\pi^2 b^4_\perp b_z^2}\left[\phi_0^4+\left(\sqrt{2}\phi_1^2-\phi_0\phi_2\right)^2+2\phi_0^2\phi_1^2\right]
\end{equation}

The corresponding NLF is displayed on Fig. \ref{fig:cdef}. A similar effect than in the case of $^{8}$Be still at work, on the surface of the nucleus, although additional oscillations appear in its center, due to the larger variety of HO wavefunction involved. The present results on $^{8}$Be and $^{20}$Ne are also in agreement with microscopic calculations (Fig. 1 of \cite{rei11}).

In summary, the large NLF value, close to the surface of the nucleus, is a general mechanism, and a necessary but not sufficient condition for alpha cluster formation. One of the 
additional conditions is of course deformation.

\section{Microscopic EDF approach }
\label{sec:micro}

\subsection{The densities and the NLF in $^{20}$Ne}

The NLF can also be computed microscopically, using EDF. The covariant EDFs are known to describe spatially localized alpha structures in light nuclei \cite{ebr12,ebr14}, visible in their nucleonic density. This is less the case with Skyrme EDF, where no such structure spontaneously appears in the ground-state density. The difference is due to the depth of the confining mean-field potential, which is larger in the covariant case, being based on the scalar and vector fields, related to the saturation properties of nuclei \cite{ebr12}. It should be noted that the covariant EDF calculations realized at the projected-GCM level, also allowed to successfully describe the rotational bands observed in $^{20}$Ne, as well as
 in $^{12}$C for the Hoyle state \cite{mar18,mar19}. In the case of Skyrme EDF, spatially localized structure in the densities can be obtained only at very large deformation of very high spin states \cite{ich11,rei11}.

\begin{figure}[tb]
\scalebox{0.35}{\includegraphics{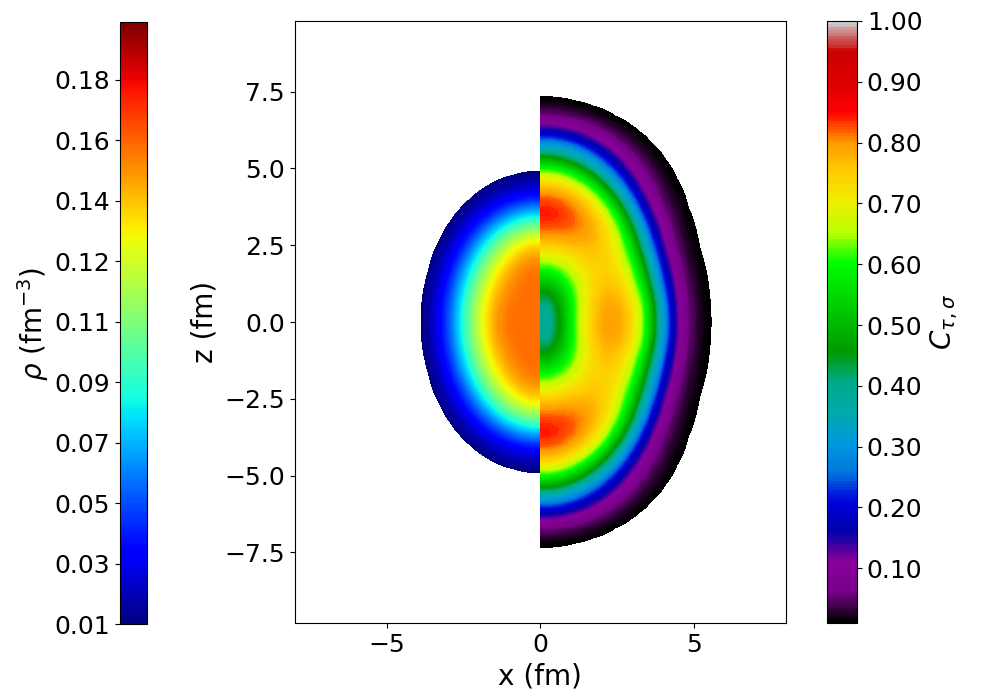}}
\scalebox{0.35}{\includegraphics{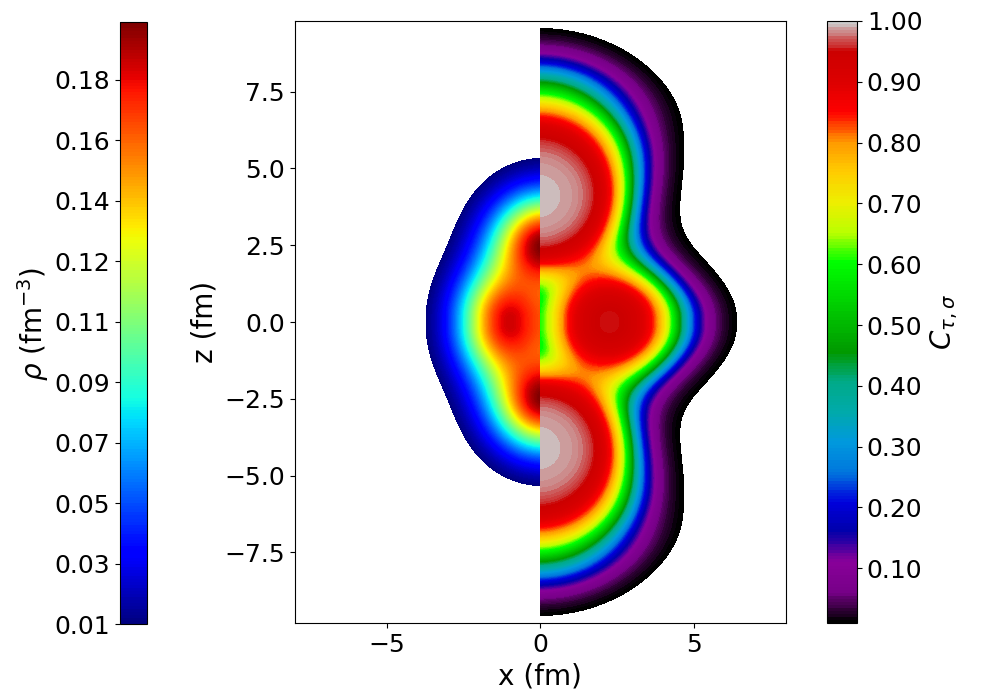}}
 \caption{Microscopic calculations of the density (left part of the figures) and NLF (right part of the figures) in the ground state of $^{20}$Ne, using the SLy5 (top) and DD-ME2 (bottom) EDF. }
 \label{fig:Skcov}
\end{figure}

Therefore, the microscopic calculation of both the NLF and the nucleonic density provides a relevant basis for understanding the relative roles of these two quantities. Fig. \ref{fig:Skcov}  displays the NLF and the density of $^{20}$Ne, obtained both with the covariant EDF, and the Skyrme one. The RMF calculations \cite{vre05} are undertaken with the DD-ME2 \cite{lal05} EDF and the Skyrme one \cite{mar22} with the SLy5 \cite{cha98} EDF. Both calculations are performed considering the quadrupole degree of freedom. The NLF are computed using Eqs. (\ref{eq:z}) and (\ref{eq:cdef}). 

Both NLF exhibit two regions of high localization. On the contrary, only the covariant density displays localized clusters, with values close to the saturation density. This confirms the previous analysis of section II, namely that the NLF is not an indicator of the spatial compactness of the alpha-particle itself, i.e. cluster formation, but rather signals
that a pure overlap of four nucleons (n$\uparrow$,n$\downarrow$,p$\uparrow$,n$\downarrow$) can be detected at a given position, irrespective of their spatial localization, or whether their form a bound state or not.

It should be noted that the position of the alpha clusters in the density and the maxima of the NLF are shifted by about 1.5 fm (bottom inset of Fig. \ref{fig:Skcov}). However, at the position of the alpha clusters, C$_{q\sigma}>$ 0.9, showing that the 4-nucleons overlap is already very pure. This means that the alpha cluster in $^{20}$Ne ground state is predicted with only a small overlap with the core of the nucleus. 1.5 fm further, the 4 nucleon overlap is completely pure on the NLF, but there is no alpha cluster on the density. This is interpreted
by a position where the 4-nucleons overlap is pure, but deals only with the tails of the wave-functions.

\subsection{Alpha-particle emission}

Recently, microscopic calculations based on covariant EDF successfully described alpha decay for several nuclei of the nuclear chart, such as $^{104}$Xe, $^{108}$Te,  $^{212}$Po or $^{224}$Ra \cite{mer20,mer21}. In this approach, a multidimensional potential energy surface is calculated considering quadrupole, octupole, and hexadecapole degrees of freedom, in order to compute the least action principle, leading to the formation and emission of the alpha-particle in these nuclei. This approach is similar to the one which has been successfully used for fission for several years (see \cite{zha19} and refs. therein). Therefore, the study of both the density and the NLF during the process of alpha radioactivity could lend relevant information on their respective role with respect to alpha-particle formation and localization.
 
\begin{figure}[tb]
    \scalebox{0.60}{\includegraphics{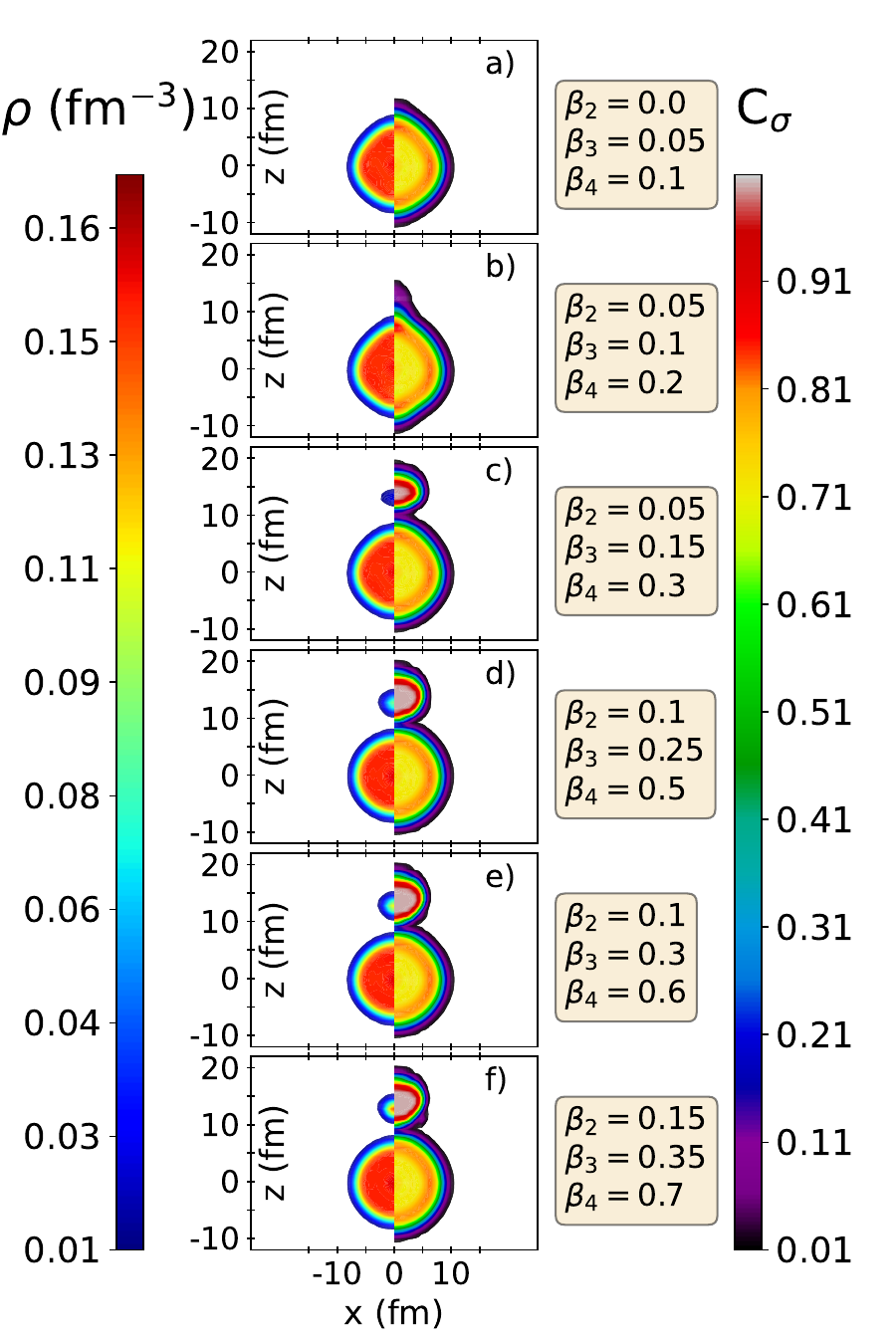}}
    \centering
    \caption{Microscopic calculations of the density (left part of the figures) and NLF (right part of the figures) during the alpha decay of $^{212}$Po, using the DD-ME2 EDF.}
    \label{fig:212Po}
\end{figure}

Fig. \ref{fig:212Po} displays these quantities, calculated in the covariant approach with the DD-ME2 \cite{nik08} EDF, during the alpha emission of $^{212}$Po. They are taken along the least action path determined in \cite{mer20}, leading to the emission of an alpha-particle by the $^{212}$Po nucleus. Let us recall, from the discussion in section II, that the NLF is sensitive to the pure spatial overlap of the nucleons (n$\uparrow$,n$\downarrow$,p$\uparrow$,n$\downarrow$), which is a necessary (but not sufficient) condition for
alpha preformation. The information of compactness and on the binding of this 4 nucleons state, is brought by the study of the density, namely localized cluster at about the saturation density. On Fig. \ref{fig:212Po}, the NLF indicates that there is a pure 4-nucleons overlap ($C_{q\sigma}=1$, in inset $c$), before the alpha-particle is fully formed in the density (inset $f$), where the number of nucleons is 4. This could be interpreted as the first step of alpha emission: the 4 nucleons first overlap in a delocalized state (inset c). This may be a necessary step of alpha-particle preformation. Later, during the emission process, the alpha-particle gets formed as a cluster (i.e. spatially localized with 4 nucleons at saturation density, inset f) in a second step. It should be noted that in this last case, the alpha cluster in the density coincides with a NLF value C$_{q\sigma}$=1, showing that it is a totally pure 
4 nucleon overlap. The present analysis enlightens the mechanism of the alpha formation process during the alpha decay: within the present approach, the alpha-particle starts to be preformed in a delocalized 4 nucleons state, and along the process, the density probability increases, so to end as a localized alpha-particle, namely a cluster, as expected in the RMF case.

It should be noted that the NLF in the middle step of the process (insets $c$ and $d$) corresponds to a similar picture to the recent one discussing the formation of alpha-particles during the fission process \cite{ren22}: NLF close to 1, with non-localized density, at value below the saturation one. This would mean that during the fission process, delocalized 4-nucleons overlap could be formed and detected, but maybe not as compact alpha clusters, as there are no alpha clusters on the predicted density profiles. It also shows the two-step process: when the nucleonic density emerges in an empty location, first the NLF can have value close to 1, meaning that a pure overlap of 4 nucleons could be detected there. In a possible, but not systematic, second step, the density increases and eventually reach the saturation density, corresponding to the presence of the 4 nucleons of a spatially localized and bound alpha-particle. But this second step may not always be reached, as indicated by the density plots of Ref. \cite{ren22}, which remain below the saturation density and show no localization at the positions where C$_{q\sigma}$=1.

\subsection{Criteria for alpha cluster formation in nuclei}

The present study allows defining criteria in order to predict the presence of an alpha cluster in nuclei. Based on the above analysis, a cluster implies that the NLF is close to one at the cluster location, and that the nuclear density is close to the saturation value predicted by the model. An additional geometric condition has to be added, to monitor the compacity of the cluster: the density should drop fast enough over the typical size of an alpha particle.
This last condition shall avoid to consider clusterisation in the ground state of $^{16}$O, despite its larges values of the NLF on the surface: as discussed above, alpha clusters cannot occur in spherically symmetric states. For this purpose, a good indicator of the density variation is the F factor, introduced in \cite{gra09} to study density variations in possible bubble candidates. It has been used subsequently and has proven to be relevant, see e.g. \cite{per22} and ref. therein. We propose here to generalize this definition by using:

\begin{equation}
F(r)=\frac{\rho_{max}-\rho(r)}{\rho_{max}}
\end{equation}

where $\rho_{max}$ is the maximum of the density of the nucleus. In the present cases, $\rho_{max}$ is usually close to the saturation density and occurs for the
center of nuclei. When F(r) is close to one, this indicates a location of low density, such as a depletion. For instance, the value of F(0) is used to monitor possible bubble effects in the center of nuclei \cite{gra09,per22}. In the present case, a value of F close to one is therefore expected at positions that delimit the alpha cluster, where the density drops. 

\begin{figure}[tb]
    \scalebox{0.25}{\includegraphics{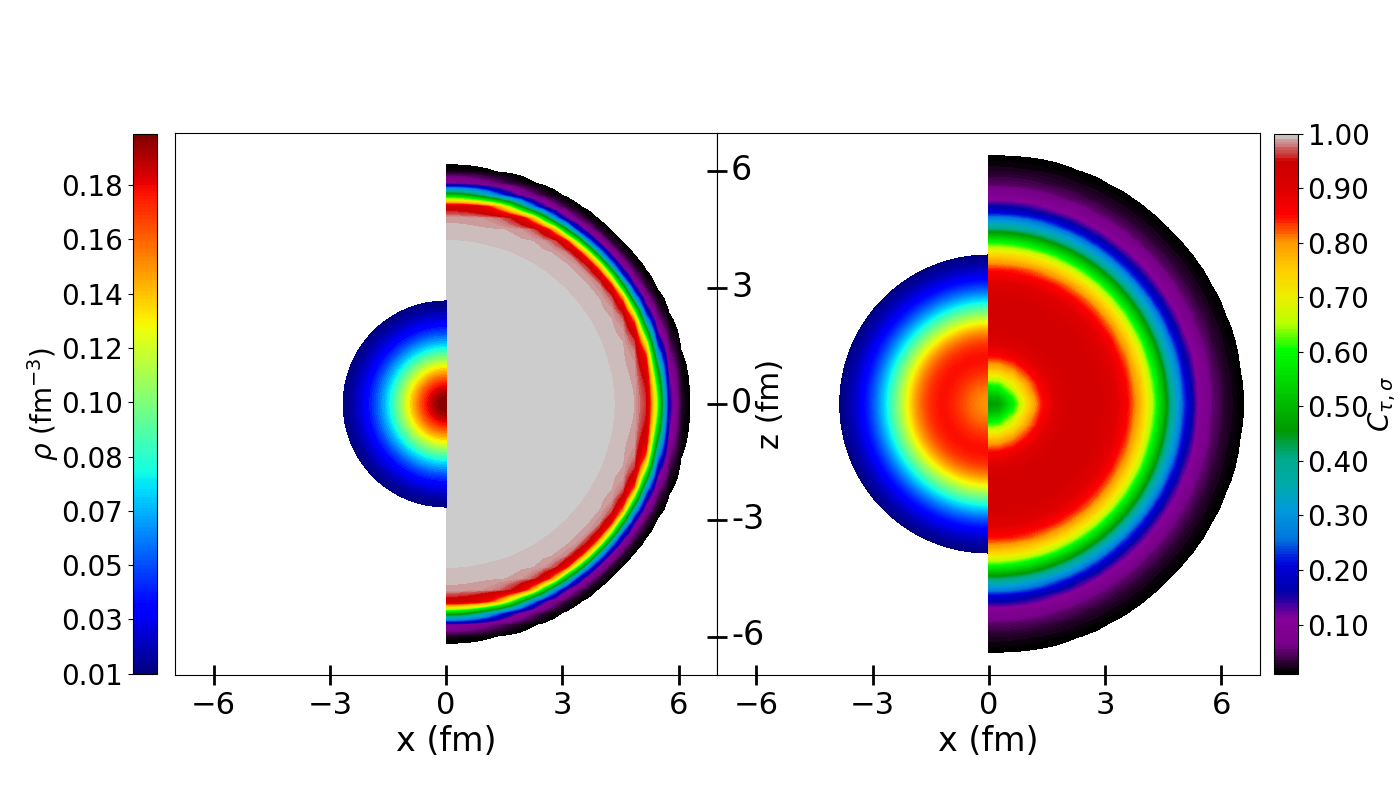}}
    \centering
    \caption{Microscopic calculations of the density (left part of the figures) and NLF (right part of the figures) for $^{4}$He (left) and $^{16}$O (right), using the DD-ME2 EDF.}
    \label{fig:4He}
\end{figure}

Fig. \ref{fig:4He} displays the density of $^{4}$He predicted with the covariant DD-ME2 EDF. Of course,  over the alpha radius R$_\alpha\simeq$ 1.9 fm, the density significantly drops. Hence, F(R$_\alpha$) close to one is a good indicator that the density drops over the typical size of an alpha particle. In a nucleus, F being close to one, at a distance R$_\alpha$ from the cluster position, shall indicate some level of compactness of the cluster.

In the case of nuclei, F should also be evaluated in two directions, to ensure its compactness. In order not to be contaminated by the core, one should choose those directions respectively opposite and orthogonal to the core. It should be noted that these relevant directions could be found by looking to increasing values of NLF (pointing towards the surface).

In summary, a sound indication for an alpha cluster prediction in a nucleus, is obtained when all the following four conditions are met, at the position where the cluster is expected:
i) C$_{q\sigma}$ $>$ 0.9, ii)  $\rho>$0.8 $\rho_{max}$, and considering the two above mentioned directions:  iii)  F(R$_\alpha$)$_\parallel>$0.6 and  iv) F(R$_\alpha$)$_\bot>$ 0.6. These threshold values for the criteria have been established considering the corresponding values for 
$^4$He, which is known to be well described by RMF calculations, with a binding energy of 27.82 MeV, to be compared to 28.51 MeV for the measured value, and charge radius of 1.98 fm, to be compared to 1.68 fm \cite{kra21}.

Table \ref{tab:cc} provides the corresponding values for these criteria in the case of the benchmark $^4$He case, and for $^{16}$O and $^{20}$Ne, calculated with the covariant DD-ME2 EDF. The values for $^{20}$Ne, calculated with the Skyrme EDF, are also displayed. The values of Table \ref{tab:cc}  are taken at the location of the maximum of the density, where an alpha cluster is suspected. Of course, the values for $^{4}$He fulfill all the criteria, as it is the benchmark case that allowed to define the corresponding threshold values for these four constraints. In the case of the DD-ME2 calculation on $^{20}$Ne, all the criteria are fulfilled as well, indicating a sound prediction for the presence of an alpha cluster. In the case of the SLy5 calculation on $^{20}$Ne, the NLF value is already too low at the suspected cluster position from the inspection of the density. Moreover, none of the F criteria are fulfilled in this case. 

\setlength{\tabcolsep}{6pt}
\begin{center}
\begin{table}[h]
\begin{tabular}{c|ccccc}
  & C$_{q\sigma}$ & $\rho$/$\rho_{max}$   &  F(R$_\alpha$)$_\parallel$ & F(R$_\alpha$)$_\bot$
 \\ \hline
 Criteria for cluster & $>$0.9   &  $>$0.8   & $>$0.6 & $>$0.6 
\\\hline
  $^{4}$He (DD-ME2) & 1   &  1   &  0.82 & 0.82 
\\ 
 $^{20}$Ne (DD-ME2) & 0.95 &  1  &  0.75 &  0.70 
\\
 $^{20}$Ne (SLy5) & 0.6 & 1 & 0.45 &  0.25 
 \\ 
 $^{16}$O (DD-ME2) & 0.9 &  1  & 0.79 & 0.41  
 \\
\end{tabular}
\caption{\label{tab:cc}}
The four criteria for alpha cluster in nuclei, and their values in  $^{4}$He, $^{16}$O, and $^{20}$Ne using the DD-ME2 functionnal in mean-field calculations. In the case of 
$^{20}$Ne, the value obtained with the Skyrme SLy5 functionnal is also displayed.
\end{table}
\end{center}
\setlength{\tabcolsep}{6pt}

Finally, in the case of the DD-ME2 prediction on the spherical $^{16}$O nucleus, both the density and the NLF criteria are met, as can be seen in Fig. \ref{fig:4He}. However, the geometrical criteria are not all fulfilled, as shown by the F$_\bot$ value on the orthogonal axis. This is due to the fact that the localized behavior in this nucleus corresponds to a ring, not a cluster. Therefore, the present 4 necessary criteria for alpha cluster occurrence in nuclei seem robust enough to take into account the various cases and detect alpha cluster predictions.

\section{Conclusion}

The respective roles of the nucleonic density and of the nucleonic localization function have been studied in the framework of alpha-particle formation and clusterization. 
The analytic derivation of the NLF, in the 3D HO approximation, shows that large values (i.e. close to 1) are obtained in the locations where the centrifugal effect vanishes or is small: in the case where only the $\ell$=0 state is filled (alpha-particle itself) or in the surface of the nucleus. In the spherical case, the NLF is driven by the ratio of the typical inter-nucleon distance to the distance to the center of the nucleus. On the surface, this ratio is small, leading to a large value of the NLF. Beyond the surface, the inter-nucleon distance suddenly increases, leading to the vanishing of the NLF far out of the nucleus. 

However, the NLF is not sensitive to the spatial compactness of the alpha-particle itself: the NLF microscopically calculated in the Skyrme EDF and in the RMF cases are similar whereas the corresponding nucleonic densities differ: delocalized in the case of the Skyrme one, and localized in the case of the RMF one. The NLF indicates the location
of pure 4-nucleons overlap, independently of their compactness. In a complementary way, the covariant density displays the locations where the alpha-particle gets localized, in a cluster state. This interpretation is confirmed by the microscopic study of alpha decay in heavy nuclei, where the NLF indicates, in a first step, the possible detection of pure 4-nucleons overlap at the surface of the nuclei, before the density shows the formation of an alpha cluster, well localized and with 4 nucleons. Based on the present analysis, four criteria could be extracted from the  NLF and the density, all to be fulfilled, in order to assert the prediction of alpha-particle clusters. 

It should be noted that a better indicator for alpha clusterization should be based on a four particles correlator measurement. Such an investigation could be undertaken in the future by considering beyond mean-field approaches.


\onecolumngrid
\appendix
\section{Derivation of a relation involving spherical harmonics \label{app}}

We derive the following relation on spherical harmonics, which we could not find in textbooks, such as \cite{abra,brin,vars,grad}.:

\begin{equation}
\sum_{m=-l}^l \mid\partial_\theta Y_l^m(\theta,\phi)\mid^2=\frac{l(l+1)(2l+1)}{8\pi}
\label{eq:dem}
\end{equation}
 
 The kinetic energy density is defined by \cite{vau72}

\begin{equation}
\tau_{q\sigma}(\vec{r})=\sum_{n,l,m}\vec{\nabla}\Phi_{n,l,m}^*(\vec{r}).\vec{\nabla}\Phi_{n,l,m}(\vec{r})
\end{equation}

Using the spherical symmetry case of Eq. (\ref{eq:wfdef}), one gets:

\begin{equation}
\tau_{q\sigma}(r)=\sum_{n,l}\frac{2l+1}{4\pi}\left(\partial_r\varphi_{nl}(r)\right)^2+\frac{\varphi_{nl}^2(r)}{r^2Sin^2\theta}\sum_{m=-l}^l m^2\mid Y_l^m(\theta,\phi)\mid^2+
\frac{\varphi_{nl}^2(r)}{r^2}\sum_{m=-l}^l \mid\partial_\theta Y_l^m(\theta,\phi)\mid^2
\label{eq:der}
\end{equation}

Inserting the following tabulated relation \cite{vars}:
\begin{equation}
\sum_{m=-l}^l m^2\mid Y_l^m(\theta,\phi)\mid^2=\frac{l(l+1)(2l+1)}{8\pi}Sin^2\theta
\label{eq:tab}
\end{equation}

 in (\ref{eq:der}), and equalizing  (\ref{eq:der}) with (\ref{eq:tau}), yields (\ref{eq:dem}).

\twocolumngrid

\end{document}